# An efficient fluorescent single-particle position tracking system for long-term pulsed measurements of nitrogen-vacancy centers in diamond

Kiho Kim, Jiwon Yun, Donghyuck Lee, and Dohun Kim*

*Department of Physics and Astronomy, and Institute of Applied Physics, Seoul National University, Seoul 08826, Korea*

*Corresponding author: dohunkim@snu.ac.kr

**Abstract**

A simple and convenient design enables real-time three-dimensional position tracking of nitrogen-vacancy (NV) centers in diamond. The system consists entirely of commercially available components (a single photon counter, a high-speed digital-to-analog converter, a phase-sensitive detector-based feedback device, and a piezo stage), eliminating the need for custom programming or rigorous optimization processes. With a large input range of counters and trackers combined with high sensitivity of single-photon counting, high-speed position tracking (upper bound recovery time of 0.9 s upon 250 nm of step-like positional shift) not only of bright ensembles but also of low-photon-collection-efficiency single to few NV centers (down to $10^3$ s$^{-1}$) is possible. The tracking requires position modulation of only 10 nm, which allows simultaneous position tracking and pulsed measurements in the long term. Therefore, this tracking system enables measuring a single spin magnetic resonance and Rabi oscillations at a very high resolution even without photon collection optimization. The system is widely applicable to various fields related to NV center quantum manipulation research such as NV optical trapping, NV tracking in fluid dynamics, and biological sensing using NV centers inside a biological cell.

**Introduction**

The nitrogen-vacancy (NV) center is a quantum coherent impurity trapped in the diamond crystal lattice. The ground state spin of the NV center can be initialized and measured using spin-dependent fluorescence and manipulated by microwaves.[1, 2] NV centers are actively studied for these properties in the field of quantum information science, including various quantum devices like quantum bits [3-5] and nanoscale sensors.[6-11]

A conventional tool for observing NV centers primarily consists of a confocal microscopy system[2] that maximizes optical contrast by accepting only the signal from the focal plane using a pinhole in front of the detector. Due to its high efficiency and high resolution, confocal microscopy has been used in various optical imaging applications. For single-particle measurements, this system should be able to track the local point of interest for a long time as most measurements on single fluorescent particles suffer from low photon collection efficiency. Hence, long-term time averaging is required to enhance the signal-to-noise ratio. This is typically done by correcting the positional error caused by environmental fluctuations such as temperature variation and acoustic vibration. For example, the local photon collection process can be periodically paused, and the local intensity maximum can be found by intermittent two-dimensional area scanning (via either raster scanning or a more efficient numerical intensity maximum finding method like simulated annealing[12]).

On the other hand, position modulation based lock-in detection relies on constantly modulating the sample position or laser spot while measuring a position error signal and employing negative feedback of the position correction information. It is widely used for applications associated with high photon emission rates in various fields such as optical tweezers[13, 14] and particle position tracking,[15-19] where the particle of interest contains many fluorescent emitters. So far, however, it has been difficult to track single fluorescent emitters like single NV centers in diamond as it is limited mainly by low photon count rates.

In this paper, we develop a highly efficient and fast real-time three-dimensional tracking system that tracks the positions of fluorescent NV centers by combining a commercial photon detector and phase-sensitive detection. This simple combination of commercial equipment enables long-term stabilized photon collection at the optical diffraction limit. To perform the experiment, we included a digital-to-analog converter and simple negative feedback system in the existing pulsed optically detected magnetic resonance (ODMR) measurement setup[20, 21]. We were able to track either an ensemble or a single NV center in nanodiamond particles or in a bulk diamond plate surface for at least several tens of hours while simultaneously measuring its quantum optical properties. Additionally, the system enables reliable measurement of time-correlated single-photon counting, as well as pulsed ODMR, even for non-photon-collection-optimized single NV centers when the photon count rate is only ~ 5000 $s^{-1}$. Due to its simplicity, high efficiency, and commercial availability, the system is expected to be widely utilized in the field of real-time position tracking with minimal external position disturbance.

**Position feedback design**

**Key components**

Figure 1(a) shows the optical schematic of the experimental apparatus. We used a continuous wave green laser (CNI, MLL-III-532-200mW) to track the position of NV centers. Simultaneously, this green laser is used for measuring the pulsed spin resonance characteristics of NV centers with an acousto-optic modulator (AOM, Crystal Technology, AOMO 3080-125), which creates a pulse of laser light with ~ 30 ns rise time. The red light emitted from an NV center is split into two paths by a single-mode fiber beam splitter (Red SM fiber BS), serving as a pinhole, and detected by two single-photon counting modules (avalanche photodiode

(APD); APD A and APD B, Exelitas, SPCM-AQRH-14-FC). This structure is identical to the Hanbury–Brown–Twiss setup that allows for second-order autocorrelation $g^{(2)}(\tau)$ measurements, where $\tau$ is the time delay between APD A and APD B. For simultaneous position tracking and spin resonance measurement, we use a power splitter to split the signal from the APD into two parts (Fig. 1(a)). Port A1 and Port B2 are connected to the apparatus for measuring the opto-spin characteristics of NV centers. Namely, they are connected either to a time-correlated single-photon counter (PicoQuant, TimeHarp 260 Nano) for $g^{(2)}$ measurements or to a high-speed photon counter (Stanford Research Systems, SR400) for pulsed ODMR measurements. Meanwhile, Port A2 and Port B1 are connected to another SR400 to track the NV center positions. The second SR400 outputs an analog voltage signal proportional to the photon count signals that are fed to both Port A2 and Port B1 using a built-in digital-to-analog (D/A) converter. The output analog voltage signal is input to a phase-sensitive detector-based position feedback device (Thorlabs, TNA001 T-Cube NanoTrak Controller) for negative feedback. The NanoTrak controller, which was originally developed for automatic alignment of fibers in conventional optical experiments, performs position modulation, phase-sensitive detection, and automatic alignment of fluorescent particles in this experiment.

Figure 1(b) shows the simplified principle and block diagram of the negative-feedback-based position stabilization. Position modulation allows for measuring the spatial derivative of the photon intensity proportional to the deviation from the fluorescent center. This error signal can be used to dynamically correct the position error with an appropriate negative feedback gain. Moreover, NanoTrak exploits circular position modulation. Thus, in-plane and quadrature signals can be detected simultaneously, proportional to the x-axis and y-axis position errors, respectively. In this work, we use two NanoTrak modules to simultaneously control and stabilize three-dimensional position errors.

More specifically, the photon pulse signal generated by the APD is input to the SR400 via Port A2 and Port B1, while the SR400 outputs an analog voltage proportional to the number of pulse signals acquired during the detection time $\tau$ (typically set to 1 ms). This analog signal is input to the NanoTrak, and the spatial derivative of the photon signal is measured. The NanoTrak modules produce three sinusoidal outputs applied to the x, y, and z inputs of the piezo stage (Thorlabs, MAX312D). First, ~ 25 Hz modulations with 90° phase shift (i.e., sine and cosine waves) are applied, which move the NV center in a circular path in the x-y plane. Second, a sine wave with ~ 35 Hz is applied for z-axis modulation. NanoTrak performs lock-in (phase sensitive) detection of the incoming light intensity oscillating at the modulation frequency by multiplying the input signal by the reference signal. The detected error signal is multiplied by a negative feedback gain (usually ×100) and used to compensate the position error.

With the minimum dwell time (D/A update time) of the SR400 of 2 ms[22] and photon detection time of 1 ms, the minimum photon count to analog voltage update time is ~ 3 ms. This update time limits the position modulation frequency, or the overall negative feedback rate, to less than ~ 50 Hz. However, it can be easily improved using, for example, field programmable gate array (FPGA) programming[23] to build a dedicated pulse counter and fast D/A converter. Other commercial equipment can also be considered, like the photon-to-voltage converter from IonOptix that has a faster analog voltage update rate but limited available output voltage range and is probably only suitable for bright ($> 10^5 \, \text{s}^{-1}$) fluorescent particle tracking.[24]

**Feedback setting details**

The position feedback system described above is assembled as shown in Fig. 1(a), with the equipment settings for efficient position feedback as follows. The feedback gain of the

NanoTrak is always set to the minimum value of ×100. The D/A output voltage range of the SR400 is set to $10^3$ counts per 1V for a dark, single NV center with several thousands s$^{-1}$ and set to $10^4$ counts per 1V or $10^5$ counts per 1V for a bright NV center ensemble (tens of thousands s$^{-1}$). In order to minimize the position perturbation due to the position modulation, the modulation radius is always set to less than 20 nm, and typically to 10 nm. The position feedback is controlled using NanoTrak's commercial control program (APT software). Furthermore, the real-time 3-axis position is obtained by measuring the monitor output voltage of the piezo stage using a National Instrument Data Acquisition card (NI 6211).

**Sample preparation**

To demonstrate position tracking, we prepared two types of NV center specimens. First, we prepared nanodiamonds dispersed on a 5 mm × 5 mm quartz substrate that was cleaned in an ultrasonic bath with acetone and isopropyl alcohol (IPA) for 5 minutes. As-received nanodiamonds with a nominal diameter of 45 nm (NaBond Technologies, China) were suspended in Milli-Q water at a concentration of $10^{-4}$ mg/mL and sonicated for 10 minutes. The position feedback was performed using NV centers in bulk diamond. Based on quantum grade type IIa diamond (Element 6), to increase the NV center concentration, ion implantation was performed at an accelerating voltage of 20 keV for ~ $5 \times 10^{10}$ cm$^{-2}$ dose. After implantation, the sample was annealed at 700 °C for 3 hours.[25, 26]

Figure 1(c) and Figure 1(d) show confocal fluorescence maps of NV centers in nanodiamonds on quartz and ion-implanted type IIa bulk diamond, respectively. In the case of nanodiamonds, most diamond particles contain ensembles of NV centers with more than $10^5$ s$^{-1}$ photon count rates. Several of them exhibit count rates on the order of $10^4$ s$^{-1}$, typical for single or few NV centers. We focus on particle tracking with photon count rates on the order of $10^4$ s$^{-1}$ for few- to single-particle control. On the other hand, the ion-implanted type IIa surface contains well isolated single NV centers with photon count rates of several thousand s$^{-1}$, suitable for demonstrating single NV center photon count stabilization, as well as simultaneous position tracking and pulsed ODMR measurements. We find that the lowest signal-to-noise ratio (SNR) of our method is 2.2 for a single NV center in bulk diamond. Here, $SNR = (I_p - I_b)/\sigma$, where $I_p$ and $I_b$ denotes the temporal average photon count received in 1 ms at the center position of NV, and background, respectively, and $\sigma$ is the temporal standard deviation of the photon count in 1 ms.[27]

**RESULTS**

**Negative feedback performance: stabilized photon counting**

First, we show that the system can reliably track few NV centers down to a single NV center in both nanodiamonds and bulk diamond. Figure 2(a) shows photon counts acquired for at least 10 hours with real-time three-dimensional position feedback for a nanodiamond (Fig. 2(a), red) containing 2–3 NV centers emitting ~$10^4$ s$^{-1}$ photons and a single NV center in type IIa bulk diamond (Fig. 2(a), black) emitting on the order of $10^3$ s$^{-1}$ photons. Due to the high photo-stability of diamond NV centers, stable photon counting is possible provided that the position drift is compensated. When measuring NV centers with relatively high intensities (13 kcps), the brightness of the target NV centers shows a standard deviation of 670 s$^{-1}$. The position drift for 30 hours is shown in Fig. 2(b) based on the monitor output of the piezo stage. The maximum drift is on the order of ± 200 nm for the x-axis direction, ± 100 nm for the y-axis direction, and ± 50 nm for the z-axis direction. Therefore, we find that the drift is significantly larger in the x-axis direction compared to the y- or z-axes. We ascribe this to the tension of a microwave cable attached in this direction and connected to the PCB (Printed Circuit Board) substrate on which the sample is mounted.

Although the drift is on a sub-micrometer scale, it can cause significant photon counting variations for the diffraction-limited fluorescent spot. Note that the photon counting presented in Fig. 2. is performed without microwave irradiation, and we expect that additional perturbation like high power microwave irradiation for ODMR measurements and Rabi oscillation measurement will cause more severe sample drift due to sample heating effect. Thus, position correction is essential to enable long-term quantum optics measurements. With negative feedback, we measure the root mean square (rms) variation of the photon count of less than 5% for at least 10–30 hours for both ensembles and single NV centers, demonstrating the efficiency of our system. In the following, we use a nanodiamond to test the position error recovery and an NV center in bulk diamond to discuss further simultaneous position feedback and quantum measurement capabilities.

**Negative feedback performance: position error recovery time**

In order to demonstrate fast nanodiamond position tracking, we measured the photon counting recovery time after inducing a step-like position shift with a piezo stage. Figure 2(c) shows the photon count rate recovery time for various amplitudes of the applied position shift in the x-, y-, and z-axes. The position of the NV ensemble contained in a nanodiamond was shifted by 250 nm, 300 nm, 350 nm, and 400 nm in the x- and y-axis directions and 0.7 μm, 0.9 μm, 1.1 μm, and 1.3 μm in the z-axis direction. The tracking system can restore the original photon collection rate within several seconds at most in all directions, showing that the system can compensate for a sudden position error and restore the piezo stage position to the intensity maximum. We note that the recovery times are significantly shorter for step-like positional shifts of less than 250 nm, and often impossible to detect with the 50 ms temporal resolution of the current photon counting measurement. For a 250 nm step-like shift, the fluorescence recovery data is fitted to a Gaussian response (inset to Fig. 2(c)) with a recovery time of $\tau_r = 0.9$ s as the best fit parameter. We estimate that the current recovery time of 0.9 s is the upper bound for our tracking system, limited by the low absolute photon emission rate from several NV centers and the photon count to analog voltage converting time of 3 ms.

**Simultaneous NV tracking and time-correlated photon counting / pulsed ODMR experiment**

We now discuss real-time position tracking and long-term quantum optics measurements performed on a single NV center in bulk type IIa diamond. We measured the photon count autocorrelation function $g^{(2)}$, which is typically done in a Hanbury–Brown–Twiss setup,[28, 29] to examine the photon emission statistics and pulsed ODMR simultaneously with tracking of the NV center position with negative feedback. The time-correlated photon counter (Picoquant, TimeHarp 260 Nano), which performs photon counting and photon arrival time tagging, is connected to two APDs through Port A1 and B2 (Fig. 1(a)). Figure 3(a) shows $g^{(2)}$ as a function of the time delay $\tau$ between the two APDs, where the typical data accumulation time to obtain clear photon statistics is ~ 10 hours, considering the low photon collection efficiency for a single NV center. With simultaneous position tracking, we clearly identify a significant antibunching dip at $\tau = 0$ with the value below 0.5 in the $g^{(2)}$ measurement for a clear single photon emission signal.[28, 29]

For continuous or pulsed spin resonance measurements, Port A1 and Port B2 outputs are connected to the high-speed gated photon counter (Stanford Research Systems, SR400), again with simultaneous position feedback using signals from Port A2 and Port B1. A microwave field is applied using a 100 μm diameter gold wire manually attached near the diamond substrate surface. When the continuous microwave is tuned to the resonance frequency of the

NV center in the absence of external magnetic field, the NV center is resonantly excited from the $|m_s = 0\rangle$ spin state to the less fluorescent $|m_s = \pm 1\rangle$ spin state, where $m_s$ denotes the spin quantum number along the NV axis. This allows for an efficient optical detection of the electron spin resonance (ESR), as depicted for a single NV center in Fig. 3(b). Figure 3(c) shows the Zeeman splitting of the ESR dip as a function of the external magnetic field applied at ~ 35° angle with the NV axis. The result shows that real-time position feedback also allows for long-term stable detection of the NV center spin resonance, which is useful when the energy level splitting of the spin needs to be examined as a function of various experimental parameters like the external magnetic field. Performing stable two-dimensional mapping, such as shown in Fig. 3(c), is particularly important for investigating spin interactions between the NV center electron spin and nuclear spin bath.[21, 30, 31]

Figure 3(d) shows the schematic pulse sequence for the Rabi oscillation measurements. We acquire an additional signal after spin initialization through Port B2 and divide the two signals for normalization. In order to acquire the initialized signal, the gate of Port B2 is shifted ~ 1 μs from the synchronized signal. With a fixed microwave frequency of 2.8 GHz at the external magnetic field $B_0 = 25$ G, we adjust the duration of the microwave pulse to measure coherent Rabi oscillations. Position tracking with a position modulation of only 10 nm allows for repeated acquisition of more than $5 \times 10^6$ samples per data point, which enables high-resolution Rabi oscillation measurements with a clear Rabi frequency of 8MHz.

**CONCLUSION**

Combining commercially available optical components allows us to construct a highly efficient and fast three-dimensional fluorescent particle position tracking system without customized programming or optimization processes. Utilizing a simple design, our system can stably track single fluorescent emitters for at least tens of hours, as demonstrated in the experiment, and, in principle, for much longer time in highly photo-stable NV centers. The tracking system is also capable of recovering its original target position in less than a second even with a step-like positional shift of hundreds of nanometers. Moreover, this system enables simultaneous position tracking and quantum optics measurements of the NV centers. In our experiments, we focused on NV centers in bulk diamond and modulated the sample position using piezo stages. However, the same principle and instruments can be used for modulating the scanning laser focal position, which is more appropriate if NV center particles inside a fluid,[15] a cell,[16-19, 32] or a potential trap [13, 14, 23] are to be examined. With these advantages, the tracking system is expected to have wide applicability in the field of NV center related quantum optics and opto-mechanic manipulations.

## Acknowledgements

This research was supported by the Priority Research Centers Program (No. 2015R1A5A1037668) and Creative Materials Discovery Program (NRF-2015M3D1A1070672) through the National Research Foundation of Korea (NRF) funded by the Ministry of Science, ICT and Future Planning.

**Figure captions**

**Figure 1. Instrument schematics and diamond sample. a.** Schematic of the position tracking system and the opto-spin characteristics measurement system. See text for detailed description. Highlighted with dashed lines, the fluorescent red light from the NV center is split into two APDs through the single-mode fiber beam splitter. The pulse signal from each APD is split into two signals via a power splitter: one is input to SR400 for tracking and the other is input to the other SR400 or TimeHarp for quantum optics measurements. **b.** Schematic of single-photon

counting combined negative feedback. In the presence of position modulation, the SR400 receives pulse signals through Port A2 and Port B1, performs photon counting in the short term, and converts its digital counts to analog signals. The NanoTrak receives this output, proportional to the position error, and gives negative feedback to the piezo stage for correcting position deviations from the center of the NV. **c-d.** Two-dimensional scan images of NV centers in nanodiamonds dispersed on a quartz substrate and type IIa bulk diamond, respectively. In **c**, the bright background is a coplanar waveguide fabricated on the quartz substrate.

**Figure 2. Negative feedback performance. a.** Fluorescence intensity measured from NV centers in nanodiamonds (red) containing 2–3 NV center ensembles and type IIa bulk diamond (black) containing single NV centers. **b.** 3D position tracked for 30 hours for NV centers in a nanodiamond. **c.** Position and fluorescence count recovery upon step-like positional errors. Arrows indicate the times when the step-like shifts are applied. A sudden positional shift of 250 nm, 300 nm, 350 nm, and 400 nm (from the first to the last arrow) is applied in the x- and y-axis while 0.7 μm, 0.9 μm, 1.1 μm, and 1.3 μm shifts are applied in the z-axis. Inset: fluorescence recovery data in the x-axis direction fitted to a Gaussian response with the best fit parameter of step recovery time of 0.9 seconds.

**Figure 3. Photon statistics and pulsed spin resonance measurements on a single NV center with simultaneous position tracking. a.** Photon count autocorrelation function $g^{(2)}(\tau)$ measured from a single NV center in bulk type IIa diamond with 10 hours of data accumulation showing a clear antibunching dip, $g^{(2)}(0) \approx 0.2$, for a single-photon source. Inset shows a schematic of the detection instrument configuration. **b.** Electron spin resonance (ESR)

spectrum of a single NV center at zero magnetic field. The upper panel shows a schematic of pulse sequence timing and detection instrument configuration. **c.** Two-dimensional map of the Zeeman splitting ESR dip in an external magnetic field acquired for 10 hours. **d.** Upper panel: schematic pulse sequence for measuring coherent Rabi oscillations. Bottom panel: coherent Rabi oscillations of a single NV center with simultaneous position tracking showing a clear optical contrast of 30% and clear Rabi frequency of 8 MHz. For each data point, the pulse sequence is repeated for $5 \times 10^6$ times to increase the signal-to-noise ratio.

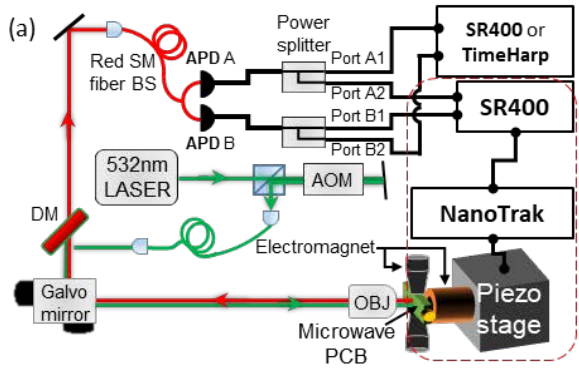
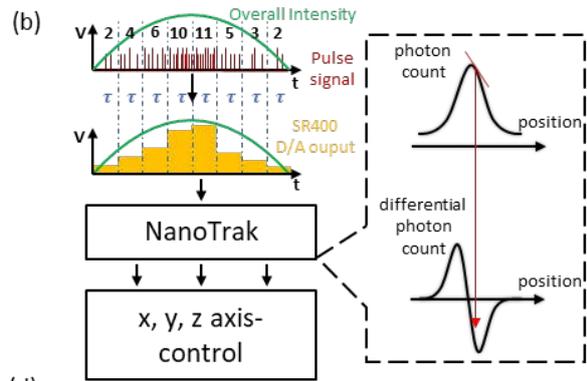
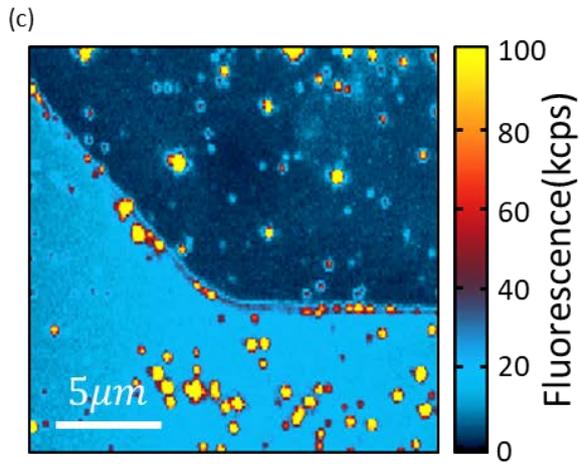
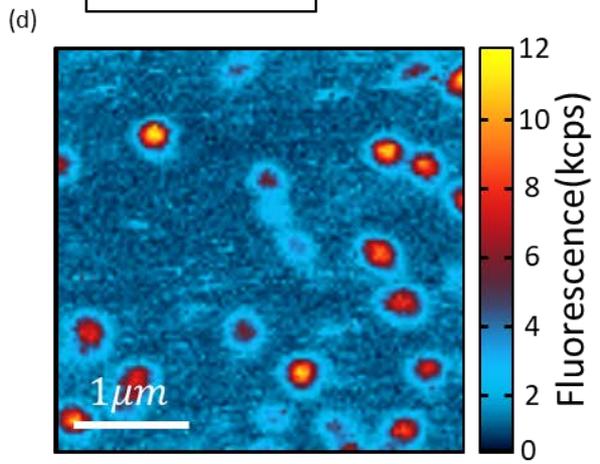

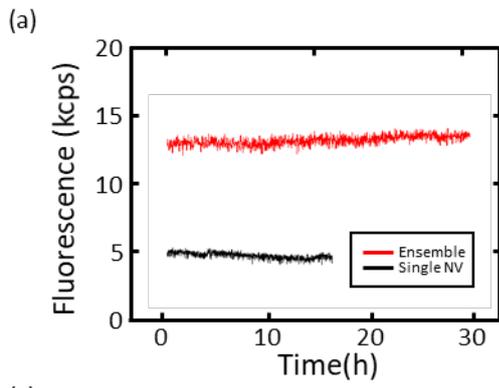
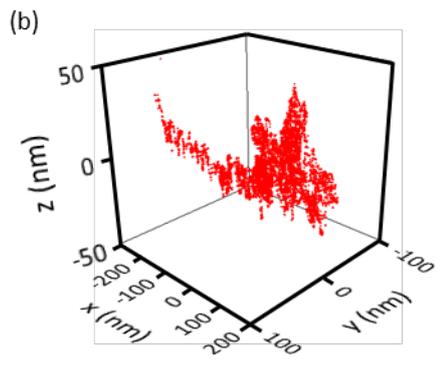
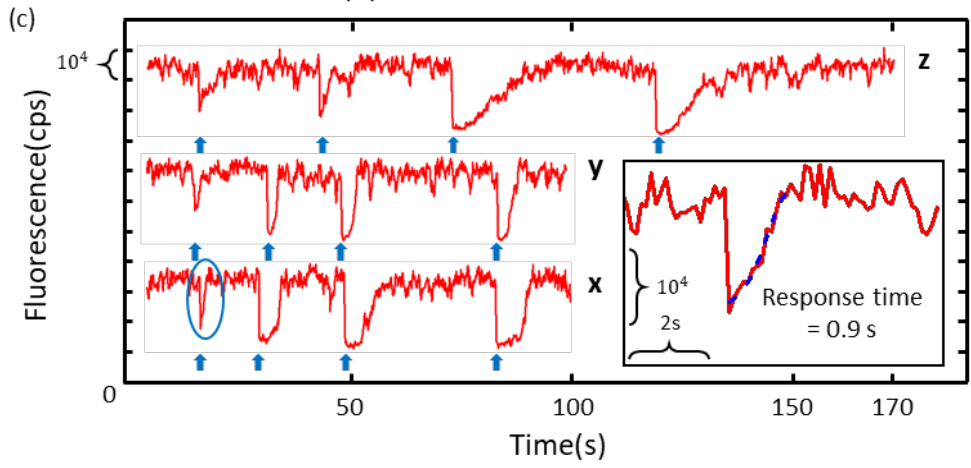

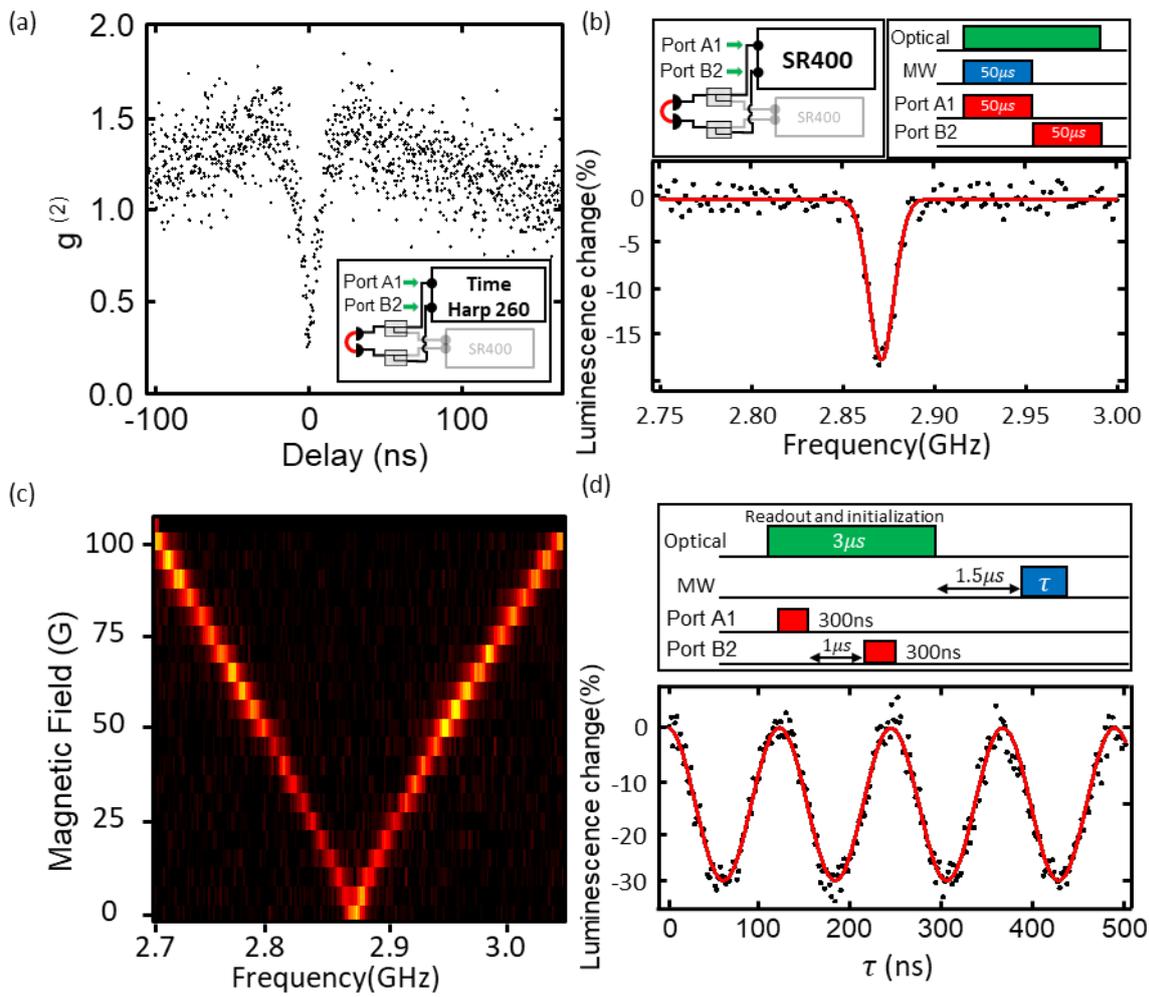